\documentclass[aps,amsmath,a4paper,twocolumn]{revtex4}
\usepackage{graphicx}
\usepackage{amsfonts}
\usepackage{amsmath}
\usepackage{ltxtable}
\usepackage{CJK}
\usepackage{mdwmath}
\usepackage{color}

\newcommand{\mG}{{\mathcal G}}
\newcommand{\mE}{{\mathcal E}}
\newcommand{\mZ}{{\mathbb Z}}
\newcommand{\mI}{{\mathbb I}}
\newcommand{\mJ}{{\mathbb J}}

\begin{document}

\title{Pushing the Qubit Rate Limit for Fast Quantum Data Transmission by Quantum Error Correction}
\author{Weidong Tang}
\affiliation{Key Laboratory of Quantum Information and Quantum Optoelectronic Devices, Shaanxi Province,
 and Department of Applied Physics
of Xi'an Jiaotong University, Xi'an 710049, P.R. China}

\author{Sixia Yu}
\affiliation{Hefei National Laboratory for Physical Sciences at Microscale and Department of Modern Physics, University of Science and Technology of China, Hefei, Anhui 230026, China}
\affiliation{Centre for Quantum Technologies, National University of Singapore, 3 Science Drive 2, Singapore 117542, Singapore}

\begin{abstract}
Fast quantum data transmission faces several shortcomings such as the indistinguishability of some partly overlapping signals, the channel noises, and so on. Based on the encoded quantum data transmission protocol, an unconventional scheme is presented to overcome the indistinguishability problem of the encoded qubits effectively, so that the upper bound for qubit transmission rate can be considerably raised compared with the un-encoded cases.
It can be regarded as a combination of an entanglement swapping technique and a quantum error-correcting scheme, which also provides us a new way of thinking for improving other technologies such as quantum communication, quantum storage technology etc. Besides, from the quantum error correction's point of view, our technique can also be regarded as a method to deal with some families of nonlocal errors, which may facilitate the research in  the relevant field.

\end{abstract}

\maketitle

\section{Introduction}

In the last decade, with the coverage of smartphone users soaring along with the thriving of numerous web communities for various purposes such as leisure, commerce, financial investment etc, the surge for the quantity of the information flux has become a severe problem, which calls for new technologies to improve the capacity of each information processing as well as to ensure the security of it. Some significant breakthroughs people envisioned are due to the applications of quantum technologies in information processing, such as quantum computating\cite{Quant Comp,Nielsen,Knill}, quantum communication\cite{Ekert,QCExp1,QCExp2}, quantum memory\cite{DLCZ,Simon,Sinclair,Tang} etc. Among these technologies or more specifically technologies for future quantum computers, to link several quantum tasks such as
quantum data producing and processing, a key step is quantum data transmission(QDT), which may dominate the speed of the relevant quantum information procedure. And many progresses of QDT have been made which include quantum teleportation\cite{Teleportation}, dense coding\cite{Nielsen,Densecoding}, key distribution\cite{Ekert,BB84,Shibata,Freespace1}, error correcting\cite{Code913,Code513,Code2} etc. Amid nearly all these technologies, entanglement plays a central role and it enables us to propose  proper schemes to overcome the limitations we faced during the information transmission.

Theoretically, any movable quantum systems can be used for QDT. But generally,
a more ideal choice is to use some bosons, e.g. photons, as they are hardly interact with each other and are easier to manipulate. Therefore, for simplicity we mainly discuss photons used for QDT in this work, the relevant techniques may also be extended to
other analogous quantum systems. However, photonic QDT may
confront several shortcomings.
Generally, besides the fidelity, there are some other significant specifications in QDT, known as the qubit rate and the transmission distance. Localized quantum tasks such as QDT during quantum computing process may not need to consider the effect caused by the distance. But some long-distance quantum tasks such as quantum communication or key distribution require the data transmit further and denser within a tolerable error threshold, spatially and temporally. The main limitation for long-distance QDT is the channel noise, which causes a severe photon loss and leads to a exponentially decay of the signal fidelity. Thus sorts of  quantum repeaters\cite{Repeater1,DLCZ} have been raised to overcome it.

Whereas in high-qubit-rate QDT, except for the technical factors such as the capacity of the computers used in the information processing etc,
just from the physical point of view, the indistinguishability of some temporal overlapping signals will be the dominant limitation eventually and it may get worse if the number of the data senders is increased. However, effective techniques to solve this problem are still very rare so far. And it may become a bottleneck for the future development of many fast quantum information tasks.

In this paper, to push the raw qubit rate limit for a QDT (un-encoded case), or in other words, to realize a higher qubit-rate QDT, we propose an unconventional scheme which is nicely immune to some overlaps for quantum signals, which can be summarized to a encoding-measurement-decoding procedure.  Our scheme can be achieved by a quantum error-correcting code (QECC) technique in which the errors are not caused by the environment but from our additional measurements, which can be regarded as a step of entanglement swapping. Following this, we give some analogous graphical QECC models to  analyze it specifically and illustrate how to push the limit of the qubit rates.
Moreover, it provides us an inherent insight for the familiar teleportation or  entanglement swapping\cite{E-swapping} technique.   Besides, we discuss how the new qubit rate upper bound depends on the temporal distributions of the physical qubits in  encoded signals via two specific models. In the end, we  also propose some potential applications.

\section{Random qubit temporal overlaps as a main limitation in fast quantum  data transmission}

\begin{figure}[tb]
\begin{center}
 \includegraphics[scale=0.7]{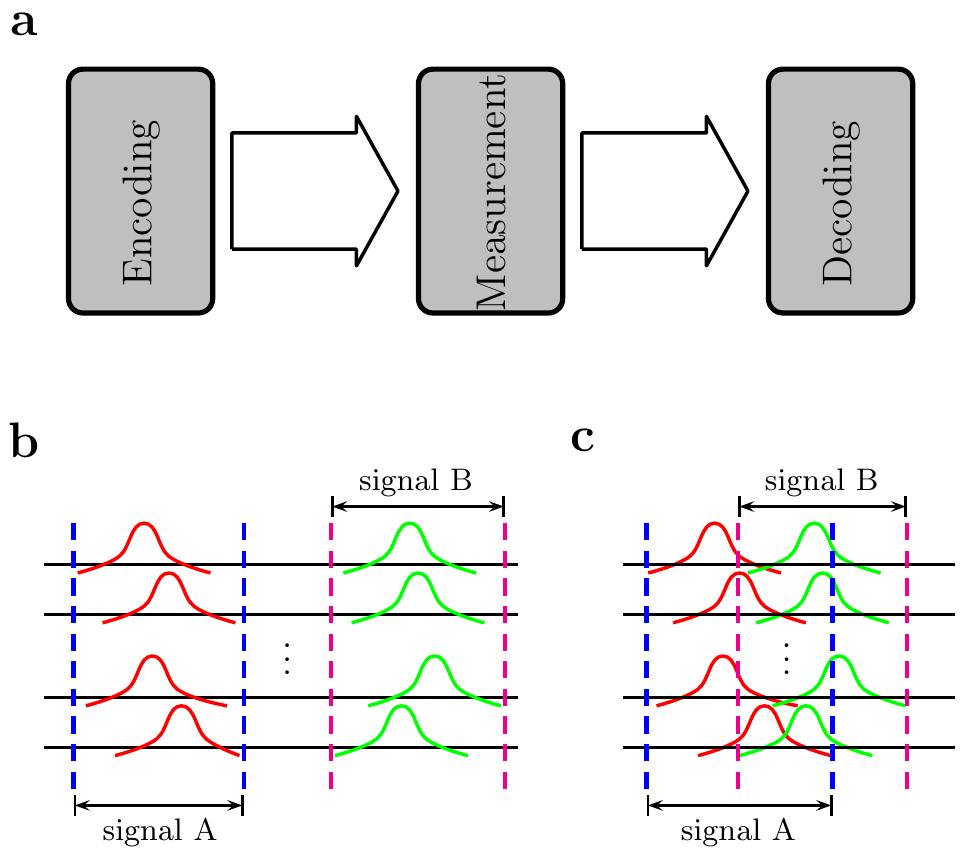}
\end{center}
\caption{\label{fig1}
{ Fast QDT protocol and one of its key restrictions --- data overlapping problems.}
({a}) An illustration of the encoding-measurement-decoding procedure for our high-qubit-rate QDT strategy.
({b}) A conceptual diagram for the  channel described in ({\bf a}) with some encoded quantum data loaded. In practice, physical qubits in each encoded signal obey a certain distribution  along the time axis .
({c}) Encoded qubits temporal overlaps are common in high-qubit-rate QDT, and here is a possible example for two sequential encoded states with part physical qubits overlapped.
}
\end{figure}

The qubit rate upper bound of QDT, which is also referred to as the qubit rate limit here, is determined by many factors. But in our model what we care about is one of the main factor --- the signals' temporal overlapping problem.  Thus the raw qubit rate limit  is inversely proportional to the effective duration of each quantum signal if it is transmitted directly without encoded to some composite system in a larger Hilbert space. Here we design an encoding-measurement-decoding technique to push the qubit rate limit of the transmitting quantum signals around a certain duration.
Before our discussion, we would like to give some assumptions to simplify our task. That is, besides the randomly signal overlapping, all other conditions are ideal.
Specifically, each physical qubit source is an ideal single photon source (or other single qubit sources);  an encoding-measurement-decoding route is introduced, i.e., each  logical qubit is encoded to a system containing several physical qubits as shown in FIG. \ref{fig1}; the gates or the detectors also work with a perfect efficiency during the data transmission;  there are no photon losses (but one encoded physical qubit may appear in the duration of another, see FIG.\ref{split}-(a)) and no environment noises etc.

As mentioned above, we first encode the logical qubits to some physical qubits when sending the data. For simplicity and without loss of generality, QECC from graph state is a typical class whose encoding scheme is preferred here. After encoding, the data(a string of logical qubits) is transmitted with encoded physical qubits at a high rate.

However, these physical qubits, even belonging to the same logical qubit, are not strictly synchronized. They may be randomly distributed in a narrow range on the time axis as shown in FIG. \ref{fig1}-(b). The logical qubits overlapping results from  the case of part physical qubits(from different logical qubits)  overlapping or even  the case of  some physical qubits distributed in a reversed time order. These two cases can be summarized to the same framework in our following processing, therefore
hereinafter they are simply referred to as the physical qubits overlapping.

Next we will introduce the main part of our scheme according to a ``Encoding-Measurement-Decoding" outline.

\section{Encoding the quantum data with graphical QECC}

Let us first recall the graphical QECC\cite{GpQECC-1} framework.  Given an $n$-qubit graph state $|G\rangle$, we denote by $I,X,Y,Z$ or $\sigma_0,\sigma_1,\sigma_2,\sigma_3$ the identity and three Pauli operators $\sigma_x,\sigma_y,\sigma_z$, respectively. For simplicity, we only consider stabilizer codes here, and non-additive codes are also consistent with our following QDT framework.
A graphical QECC $[[G,K,d]]$, sometimes also written as $[[G,k,d]]$ $(k=\log_2K)$ or $[[n,k,d]]_2$, is a coding space spanned by $\{Z^{\textbf{a}^i}|G\rangle|\textbf{a}^0=(0,0,...,0),i\in\mZ_K\}$ satisfying
$\langle G|Z^{\textbf{a}^i}\mE_dZ^{\textbf{a}^j}|G\rangle=f(\mE_d)\delta_{ij}$ for any $i,j\leq K-1$, where $Z^{\textbf{a}^i}\equiv\prod_{k=1}^{n}Z^{a^i_k}_k$, $\mZ_K\equiv\{0,1,...,K-1\}$ represents the set of integers modulo $K$,   and $a^i_k\in\mZ_2$. Here $\mE_d$ represents all the error $\mE_d=X^\textbf{s}Z^\textbf{t}$
that acts nontrivially on a number of qubits that is less than
$d$, i.e., $|\sup(\textbf{s})\cup \sup(\textbf{t})|<d$,  where we have denoted by $\sup(\textbf{s})=\{j\in\{1,2,...,n\}|s_j\neq 0\}$ the support
of a $n$-component vector $\textbf{s}$. And $f(\mE_d)\in\{\pm1,\pm i\}$ if $\mE_d|G\rangle=f(\mE_d)|G\rangle$, otherwise $f(\mE_d)=0$. Notice that any state $Z^{\textbf{a}^i}|G\rangle$ is local unitary  equivalent to the graph state $|G\rangle$ .

Now we consider the data transmission process.  At the sender's side, a logical qubit is encoded to a state in an $n$-qubit graphical QECC coding space $[[G,K,d]]$ at the time $t$ and denoted by $|G^t\rangle$.  And at the receiver's side, before decoding procedure,  a stabilizer measurement, which is also called a Bell measurement hereinafter, can be performed. Namely, the set of observables to be measured are the stabilizers of the coding space and it can be denoted by $\{\mG_{\mu}|\mu={1,2,...,n-\log_2K}\}$. But before this measurement, we should make sure that the encoded physical qubits for any logical qubit are synchronized temporally. That is, if there are overlapping qubits from two encoded graph states, we should split them at first. But this operation may lead to a possibility of exchanging the temporal ordering of them. As we will see later,
exchanging two physical qubits, which is essentially equivalent to a kind of nonlocal error, can be referred to as {\it qubits distinguishing error}(QDE).
If there are no QDEs, by the definition of stabilizers for the coding space, we have $\mG_{\mu}|G^t\rangle$=$|G^t\rangle$.

For the high-qubit-rate information transmission, random QDEs are inevitable (see FIG. \ref{fig1}-(c)). For any physical qubit from an encoded state, the total probability of being involved in a QDE can be written as $p$. We first consider the case when $p$ is a small quantity to verify the feasibility of our scheme in principle, namely, the case for a QDE appearing mainly on two encoded states which are very close to each other on the time axis. In other words, for a sequence of encoded states, taking an arbitrary one for consideration,  the probability for two and more QDEs happening on it and its nearby states, written as $\sim O(p^2)$, is insignificant
(under a certain threshold).
In the end we also generalize our scheme to deal with the case when $p$ is not small enough, i.e., the high order effects with probabilities $\sim p^2, p^3$ etc are not negligible.

\section{Bell measurements to turn the QDEs to local errors}

\begin{figure*}[tb]
\begin{center}
 \includegraphics[scale=0.85]{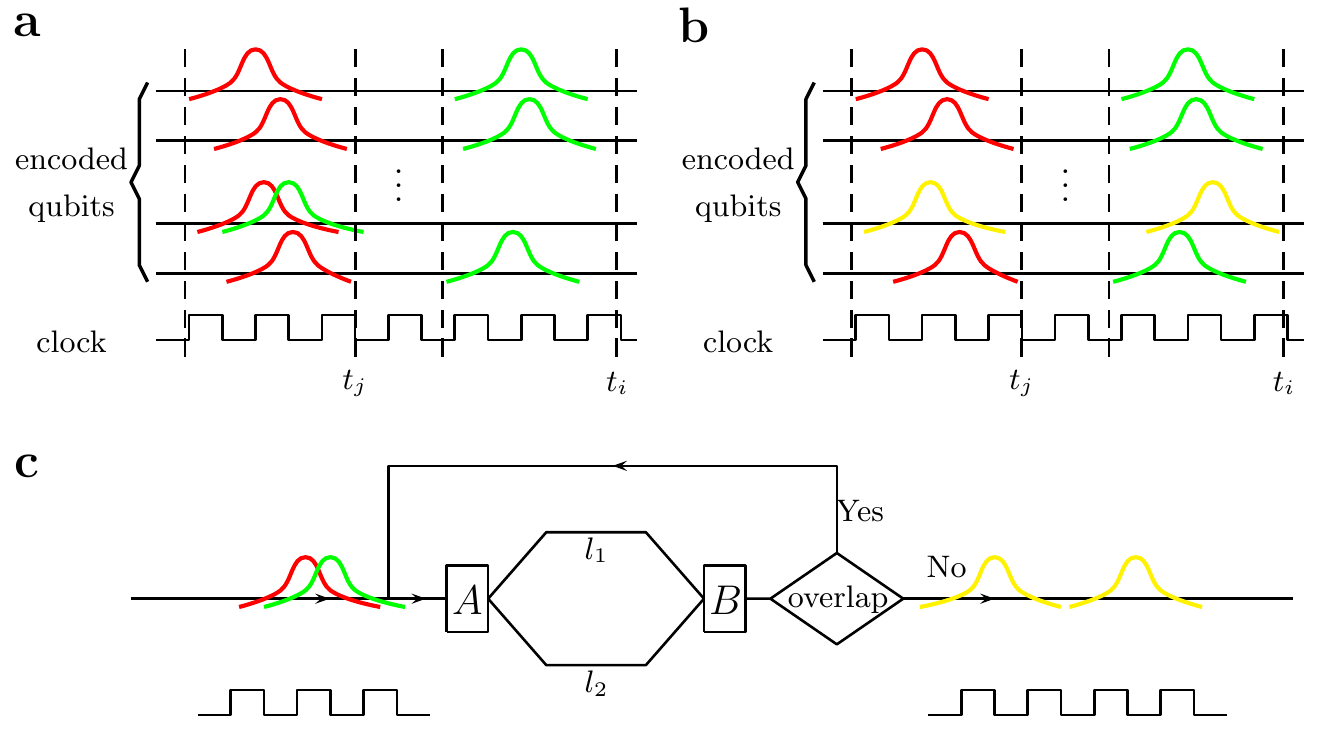}
\end{center}
\caption{\label{split}
({a})Two physical qubits from different encoded logical qubits labeled by $t_i$ and $t_j$ (or colored by green and red) respectively are randomly overlapped.
({b}) The expected physical qubit distribution after splitting, which seems approximately synchronized. Qubits colored by yellow is due to the possible QDE.
({c})A scheme for splitting two overlapping physical qubits. $A$ is an apparatus that sends  the input wavepackets (physical qubits)  randomly to two paths with different length $l_1$ or $l_2$. $B$ is used for merge the two paths to one channel. After splitting, to make the two qubits synchronized with $t_i$ and $t_j$, some classical signal (clock signal etc) assisted  time control techniques which can be realized by specifically designed circuits should be used.
}
\end{figure*}

Before  performing the Bell measurements to these encoded graph states, we should first ensure the synchronization of the encoded physical qubits from each logical qubit.
We can add some classical clock signal controlled  detecting apparatuses to measure the loss of a physical qubit or the overlap of  two or more qubits, see FIG.\ref{split}-(a).  If all the encoded physical qubits corresponding to the same encoded graph state have already been nicely synchronized (no loss or no overlap) with time, which can also be referred to as normal encoded state, then we send them directly to the subsequent Bell measurement devices. Otherwise, we should take out these abnormal encoded graph states to a temporary channel (via some storage technique or a lengthened channel, e.g. extended Optical fiber for photons), to wait for the related overlapped qubits to undergo the splitting operations. As we mentioned above,  the loss and the extra gain of a qubit happen on two nearby encoded states(FIG.\ref{split}-(a)). Hence we can easily locking them by the detecting apparatuses and match them by computers.
Here we give a practical scheme to perform the splitting operation, which is illustrated in FIG.\ref{split}-(c). The technical details are not concerned in this paper. It is obvious that the splitting operations will lead to QDEs with a certain possibility. Therefore, after that,  by the specifically designed control circuits, we can seemingly synchronize (may cause QDEs) the two abnormal encoded states(FIG.\ref{split}-(b)).

Some remarks are in order. First, to avoid  disturbance to the encoded states or the non-commuting problems with the subsequent Bell measurement, the detection for the abnormal encoded graph states can be realized by using some other degree of freedom. For example, if we encoded the qubits with the polarization, we can detect the qubits with the frequency. But we should carefully control the precision for the frequency measurement, to make the disturbance to the time domain information within a acceptable range according to the uncertainty principle.
Second, the synchronization operations will be contained in the ``measurement" step shown in FIG.\ref{fig1}-(a).

Next we will see how to convert the QDEs to local errors by the Bell measurements.

For any two $d$ dimensional state $|\phi\rangle$ and $|\psi\rangle$, if an operator $V$ satisfies $V|\phi\rangle|\psi\rangle=|\psi\rangle|\phi\rangle$, then we call it {\it states exchange operator}. It can be written as $V=\Phi^{T_2}$ \cite{Werner}, where $\Phi=d|\Phi\rangle\langle\Phi|$, $|\Phi\rangle=\frac{1}{\sqrt{d}}\sum_{k=0}^{d-1}|kk\rangle$ and $T_2$ represents the partial transposition operation for the second particle. And for the two-dimensional case, states exchange operator can be represented as a symmetric form
\begin{equation}\label{V2}
V_{12}=\frac{1}{2}(I_1I_2+X_1X_2+Y_1Y_2+Z_1Z_2).
\end{equation}
We call $V_{12}$ in Eq. (\ref{V2}) as {\it qubits exchange operator}. Clearly, $V_{12}^2=I_{4\times4}$. Likewise, this definition can also be extended to {\it qubits permutation operator} which can be built from a series of  qubits exchange operators. For example, an $n$ qubits permutation operator can be written as $V_{12...n}=V_{12}V_{13}...V_{1n}$.

For a system composed of a set of qubits labeled by $1,2,...,m$, if we perform some measurements on it after exchanging the qubits $k$ and $l$ ($1\leq i,j\leq m$ and $k\neq l$), we will see a pair of ``twined errors" occurring on the two qubits, namely, if a $\sigma_{\alpha}$ $(\alpha\in\mZ_4)$ type of error is measured on the qubit $i$, then the same type of error occurs on the qubit $j$, and vice versa. This rule is regarded as {\it qubits exchange measurements statistics} (QEMS).  However, not all the errors in any case can be well distinguished after a measurement. Some errors may be degenerated due to the speciality of a system or the pre- knowledge of a system, e.g., $XX$ type and $ZZ$ type of errors acting on two qubits of a Bell state $\frac{1}{\sqrt{2}}(|00\rangle+|11\rangle)$  can be regarded as no errors happening on them. If we want to identify the twined errors faithfully, we should ensure that  all the $\sigma_{\alpha}\sigma_{\alpha}(\alpha\in\mZ_4)$ types of errors can map the system to the states which are mutually orthogonal.

Come back to our high-qubit-rate QDT process. Let us consider a general case that a pair of overlapping qubits $i$ and $j$ which are contained respectively in two systems. Without loss of generality, we  denote by $|G\rangle_{\mI}$ and $|G^{\prime}\rangle_{\mJ}$ the two graph states with qubits index sets $\mI,\mJ$ and $i\in\mI,j\in\mJ$ the two overlapping qubits. Assuming $(\sigma_{\alpha})_i(\alpha\in\mZ_4)$ or $(\sigma_{\beta})_j(\beta\in\mZ_4)$ can map $|G\rangle_{\mI}$ or $|G^{\prime}\rangle_{\mJ}$ to four mutually orthogonal states which can be distinguished by a projective measurement(Bell measurement). Then  separating the two graph states by the above splitting operation may lead to one of the following two cases with a certain odds, namely, separating $|G\rangle_{\mI}$ and $|G^{\prime}\rangle_{\mJ}$ faithfully or incorrectly distinguishing a pair of qubits between them (QDE). A QDE on $|G\rangle_{\mI}$ and $|G^{\prime}\rangle_{\mJ}$  can be equivalent to map the two graph states to two well separated ones with one of the ``twined errors" on each if we perform a projective measurement on them. This can be easily seen from
\begin{align}\label{QEMSof2}
|G\rangle_{\mI}|G^{\prime}\rangle_{\mJ}=&V_{ij}|G\rangle_{\mI^{\prime}}|G^{\prime}\rangle_{\mJ^{\prime}}\\ \nonumber
=&\frac{1}{2}(|G\rangle_{\mI^{\prime}}|G^{\prime}\rangle_{\mJ^{\prime}}+X_j|G\rangle_{\mI^{\prime}}X_i|G^{\prime}\rangle_{\mJ^{\prime}} \\ \nonumber
+&Y_j|G\rangle_{\mI^{\prime}}Y_i|G^{\prime}\rangle_{\mJ^{\prime}}+Z_j|G\rangle_{\mI^{\prime}}Z_i|G^{\prime}\rangle_{\mJ^{\prime}}),
\end{align}
where $\mI^{\prime}=\mI\bigtriangleup\{i,j\}$, $\mJ^{\prime}=\mJ\bigtriangleup\{i,j\}$ and $A\bigtriangleup B\equiv A\cup B-A\cap B$ denotes the symmetric difference of two sets $A$ and $B$. If we treat a quantum state only from the information point of view, then $|G\rangle_{\mI}|G^{\prime}\rangle_{\mJ}$ and $|G\rangle_{\mI^{\prime}}|G^{\prime}\rangle_{\mJ^{\prime}}$ are  equivalent, as well as
$X_i|G\rangle_{\mI}X_j|G^{\prime}\rangle_{\mJ}$ and $X_j|G\rangle_{\mI^{\prime}}X_i|G^{\prime}\rangle_{\mJ^{\prime}}$ etc. Therefore,  whether a QDE happens or not, after a Bell measurement,  the two graph states can be equivalently projected to one of the following four states
\begin{align*}
|G\rangle_{\mI}|G^{\prime}\rangle_{\mJ},
X_i|G\rangle_{\mI}X_j|G^{\prime}\rangle_{\mJ},
Y_i|G\rangle_{\mI}Y_j|G^{\prime}\rangle_{\mJ},
Z_i|G\rangle_{\mI}Z_j|G^{\prime}\rangle_{\mJ}.
\end{align*}
Namely, the errors on two graph states come up in pairs. And for each graph state, a random QDE referred above can be converted to no errors or a single Pauli error on the corresponding qubit. Thus whether a QDE occurs or not,  one of the encoded states $|G\rangle_{\mI}$, will turn to one of the four states in the set
\begin{equation}\label{subsystemerror}
\{|G\rangle_{\mI}, X_i|G\rangle_{\mI},Y_i|G\rangle_{\mI},Z_i|G\rangle_{\mI}\}.
\end{equation}
And the situation for $|G^{\prime}\rangle_{\mJ}$ is analogous.
Some examples given in Appendix A not only help us to understand QEMS  specifically, but also  provide us another understanding for quantum teleportation or entanglement swapping technique from our QEMS framework.

Therefore, QEMS allows us to turn a QDE to a pair of identical errors on the corresponding qubits by a Bell measurement. This provides us a nice thought to deal with the qubits overlapping difficulties in high-qubit-rate QDT.

Let us return back to our $n$-qubit graphically encoded QDT process with high rates cases. As we have assumed that $p$ is small, then let us consider the encoded state $|G^t\rangle$ if a parity measurement (a special Bell measurement) strategy for the stabilizers of the QECC $[[G,K,d]]$'s coding space is used. From the previous discussion, after the measurement, it will turn to a state in the set
\begin{equation}\label{GPQECCerror}
    \{(\sigma_{\alpha})_i|G^t\rangle|\alpha\in\mZ_4,i=1,2,...,n\},
\end{equation}
where $i$ denotes the $i$-th qubit of the encoded graph state.  
The parity measurements performed on each encoded state by the receiver can be realized by a collection of projective measurements $\{M_{\mu,\tau}\}$, where
\begin{align}{\label{paritymeasure}}
 M_{\mu,\tau}=\frac{1+(-1)^{\tau}\mG_{\mu}}{2},\mu\in\{1,2,...,n-\log_2K\}, \tau\in\mZ_2.
\end{align}
As the errors here can only be a form of $(\sigma_{\alpha})_i$,  then we have $\langle G^t|(\sigma_{\alpha})_i|G^t\rangle=f((\sigma_{\alpha})_i)$ and $\langle G^t|(\sigma_{\alpha})_i(\sigma_{\beta})_j|G^t\rangle=f((\sigma_{\alpha})_i(\sigma_{\beta})_j)$. Therefore, to ensure the orthogonality of the states in the set shown in Eq. (\ref{GPQECCerror}), we only need to choose a QECC of a distance $d\geq3$. And if we only consider the case for one logical qubit encoded to one graph state, we can choose $k=1$ to reduce $n$.
For example, we can choose the graphical code $[[5,1,3]]_2$\cite{Code513} for its optimal number of qubits used. However,  by considering the difficulties of preparing, the Shor code $[[9,1,3]]_2$\cite{Code913} from a graph state composed of three GHZ state might be a better choice in practice.

\section{Decoding the information}

No matter what code we choose,  the parity measurements in Eq.(\ref{paritymeasure}) can nicely distinguish any single error. Analogous to the error-correcting process, after the errors artificially produced by our ``measurement" step, the decoding system  will recover the logical qubits at the receiver's side.
 
We can design two independent decoding circuits in the decoding system. One is for the normal encoded graph states, and the other is for the states split from the abnormal states. By the clock labeled information, we can extract the  quantum data correctly.

Let recall the heart of the scheme for the case when $p$ is small. Each random QDE may occur after the splitting process, then it can be converted to a pair of single errors on each encoded state by a Bell measurement,  and at last these single errors can be corrected by our QECC. 
And the difference between our scheme and QECC is that here we artificially produce errors during our measurement step, while in a conventional QECC process the errors are caused by noises from the environment.
 
After the analysis in the next section, we will see that this scheme can considerably increase the rate for the QDT.

\section{Rate upper bound evaluation}

\subsection{One QDE correctable case}

\begin{figure}
\begin{center}
 \includegraphics[scale=0.85]{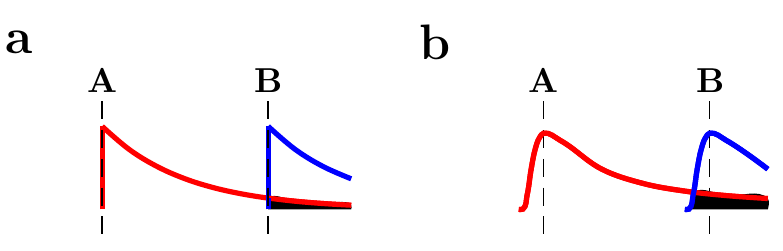}
\end{center}
\caption{\label{fig3}
{Two approximate models of qubit rate upper bound evaluation.}
({a})The exponential distribution  model. The time distribution for the physical qubits  from each encoded graph state can be described by a  exponential probability density function(note that  this differs from the waveforms for a physical qubit). The probability of the QDEs occurring  is  proportional to the overlap (the black tail zone) of two adjacent exponential distributions A and B.
({b})L\'{e}vy distribution model. Likewise, the physical qubits from each encoded signal obey a L\'{e}vy distribution temporally.
}
\end{figure}

Under the same conditions for information transmission as shown in FIG. \ref{fig1}, e.g. the same type of photons and channels, for the un-encoded case(1 physical qubit$=$ 1 logical qubit), when the  qubits are conveyed on a certain channel, there is a tolerable rate limit, assumed as $N_m$ qubits/s, to ensure the total probability of QDEs occurring on one qubit to a level under $q$ (usually $q$ is pre-given and $q\ll1$).
To give an approximate evaluation for the improvement of distinguishability for two signals in our scheme, i.e., to show how the rate upper bound $N_m$ is raised,  we consider a exponential distribution  model (EDM) and a L\'{e}vy distribution model (LDM) which are shown in FIG. \ref{fig3}.
Note that the time distribution of the physical qubits in each logical qubit is device-dependent. That is, even the physical qubits from each encoded graph state distributed in a narrow interval around sometime can be estimated initially,  they may still be affected by the encoded circuits and the gates etc and could give rise to some changes of the relevant distribution. And for the EDM(or LDM), the physical qubits  from each encoded graph state obey a exponential distribution (or a L\'{e}vy distribution) on the time axis.
The relevant probability density functions of EDM and LDM can be described by
$f_i^{EDM}(t)=ae^{-a(t-t_i)}$ and $f_i^{LDM}(t)=\sqrt{\frac{c}{2\pi}}\frac{e^{-\frac{c}{2(t-t_i)}}}{(t-t_i)^{\frac{3}{2}}}$ respectively, where $t>t_i,~ a>0,~c>0$, and $i$ labels the $i$-th encoded states of a signal string.
Herein, to show the feasibility of our scheme, the detailed analysis is only focused on EDM. But some better results are from LDM, one can see them in TABLE \ref{Table1}. And the relevant calculations can be found in Appendix B.

\begin{table*}[tbh]
\begin{tabular}{lll}
  \hline
   & EDM & LDM \\
     \hline
  $N^{\prime}_m$ ($1$ QDE's case) &
   $\frac{2N_m}{1+\frac{\log_2({3C_n^2}/{8})}{\log_2({1}/{(2q)})}} \sim2N_m$~~~& $\frac{4}{3}(C_5^2)^{-1}q^{-1}N_m$ ~~$(\sim13.3N_m;~\text{for}~[[5,1,3]]_2,~q=0.01)$ \\
  $N^{\prime}_m$ ($2$ QDEs' case) &
    $\frac{3N_m}{1+\frac{\log_2((3/8)^2C_n^3)}{\log_2{(1/(2q))}}}\sim3N_m$~~~& $(\frac{4}{3})^{\frac{4}{3}}(C_n^3)^{-\frac{2}{3}}q^{-\frac{4}{3}}N_m$~~$(\sim146.8N_m;~\text{for}~[[5,1,3]]_2,~q=0.01)$\\
  $N^{\prime}_m$ ($r$ QDEs' case)~~~ &$\sim(r+1)N_m$,~$r\leq3$~~~ & $(\frac{4}{3})^2(\frac{3}{4})^{\frac{2}{r+1}}(C_n^{r+1})^{-\frac{2}{r+1}}q^{{\frac{2}{r+1}}}q^{-2}N_m$\\
  \hline
\end{tabular}
\caption{\label{Table1}
{\bf Two approximate models of qubit rate upper bound evaluation.}
({\bf a})The exponential distribution  model (EDM), the constraints for the relevant $q$ can be found in the Results.
({\bf b})L\'{e}vy distribution model (LDM).
}
\end{table*}

For EDM's case, let $L=\frac{1}{N_m}$ (s) be the minimal average time interval for two adjacent signals corresponding to the rate of $N_m$ (qubits/s).  If the average time interval between two encoded states A and B is $l=|t_A-t_B|\geq0$, then for one physical qubit, the probabilities of QDEs occurring on it will be  $p=\frac{1}{2}(2q)^{\frac{l}{L}}=\frac{1}{2}(2q)^{N_ml}$, and the details are given in Appendix B. By our scheme,  a QDE happening on one pair of qubits from two encoded states can be corrected. As $p$ is assumed to be small in this single QDE correctable model's framework,  then we only need to consider the contributions for QDEs up to order $p^2$, i.e., keeping the odds of encoded states unrecoverable  due to the QDEs appearing on two pairs of qubits\cite{highorder}, which can be written as $(\frac{3}{4})^2C_n^2p^2(1-p)^{n-2}\approx(\frac{3}{4})^2C_n^2p^2$ (as long as $n$ is not too large) corresponding to the QECC $[[n,1,d]]_2$, less than or equal to $\frac{3}{4}q$ (compared with the un-encoded case).  Then we have $(\frac{3}{4})^2C_n^2(\frac{1}{2}(2q)^{N_ml})^2\leq \frac{3q}{4}$, where the factor $\frac{3}{4}$ referred above comes from the fact that when a QDE occurs, there is still a $25\%$'s chance to get the correct information from the expression of qubits exchange operator. Thus an approximate relationship which reads $l\geq\frac{1}{2N_m}(1+\frac{\log_2({3C_n^2}/{8})}{\log_2({1}/{(2q)})})$ is obtained. Usually $q$ satisfies $q<\frac{4}{3C_n^2}=\frac{8}{3n(n-1)}$ as long as $n$ is not too large, such as $5$ and $9$ in two examples of QECC given above, then $\frac{1}{2N_m}(1+{\log_2({3C_n^2}/{8})}/{\log_2({1}/{(2q)})})<L$. And the maximal rate can be improved to  $N^{\prime}_m=\frac{1}{l_{\min}}=\frac{2}{1+\frac{\log_2({3C_n^2}/{8})}{\log_2({1}/{(2q)})}}\cdot N_m>N_m$.
For the extreme case that $q\ll\frac{8}{3n(n-1)}$, the rate trends to $2N_m$, which gives a improvement of nearly $100\%$. The calculation for the LDM is analogous. Generally, the rate upper bound is model-dependent, which can be seen clearly from TABLE \ref{Table1}.

\subsection{Two or more QDEs correctable cases}

The above approach can be generalized and the splitting operation in the ``measurement" step should be redesigned as there may be contributions by three or more qubits overlaps. Next let us check the case when $p$ is small but not small enough to ignore some of its higher order effects. And no doubt that the number of the QDEs rises. We can also convert them to several Pauli errors even there are qubits permutations.  For example, we consider the case $d=5$, i.e., the QECC can detected $4$ qubits errors or correct $2$ qubits errors. It is immune to  QDEs happening on two or less  qubits for each encoded state. Likewise, the main limitation for separating the encoded states comes from the case that three qubits from one state are involved in the QDEs. Therefore,  the approximate qubit rate upper bound can be written as $N^{\prime}_m=\frac{3N_m}{1+\frac{\log_2((3/8)^2C_n^3)}{\log_2{(1/(2q))}}}$ and it tends to $3N_m$ for the extreme case of $q\ll\frac{64}{3n(n-1)(n-2)}$ which implies a nearly $200\%$ 's improvement. Generally, suppose we choose a QECC of $[[n,1,2r+1]]_2$ with a small $r$(e.g., $r\leq3$), and if $q$ satisfies
\begin{equation}\label{qc}
q\ll \frac{1}{2}\cdot(\frac{8}{3})^{r}\cdot\frac{(r+1)!}{n(n-1)...(n-r)},
\end{equation}
the upper bound of the rate approaches $(r+1)N_m$ when $p\sim \frac{1}{2}(2q)^{\frac{1}{r+1}}$ remains small and $(1-p)^{n-r-1}\sim1$, namely, the rate limit can still be considerably improved. But for a general case, the qubit rate limit increases nonlinearly with $r$, and it is smaller than $(r+1)N_m$. One can see that from the Appendix C.
Note that $q$ is  held fixed rather than varied, thus we can choose a suitable QECC according to the code distance. In general, for a certain model,  the larger the code distance for a QECC of type $[[n,1,d]]_2$ is chosen,  the greater the qubit rate upper bound may be obtained.

Although different models of physical qubits distributions may be used, the general physical results are similar --- the qubit rate upper bound can be remarkably raised. Some model may even give a improvement by two orders of magnitude such as LDM. But how much can the rate be raised mainly depends on the relevant distribution, and to be specific, the asymptotic tail behavior of the probability density function. Generally, the long tail's case is better than the short tail's case, e.g. LDM versus EDM. More interestingly, from the resource point of view, there may be a ``$1+1>2$" effect for some distributions. E.g., for the LDM and the code $[[5,1,3]]_2$ in TABLE \ref{Table1}, compared with the un-encoded case, we use only five qubits to raise the QDT rate by more than ten times.

\section{Discussion}

In our scheme, to fight the QDEs in quantum information transmission with high rates, a natural thought is trying to turn them to some local Pauli errors and then correct them. Therefore, at first sight,  the QECC framework may be very useful, and we try to encode the data to the coding space of some QECC.  Further, inspired by quantum teleportation and entanglement swapping techniques, we realize the above thought by a measurement process and a particular decoding procedure. Finally we get the corrected quantum data at the output of the decoding device. This is the full ``encoding-measurement-decoding" process. By using different QECC, we can in principle push the raw qubit rate limit for the QDT,  and may raise it by several times or even one to two orders of magnitude compared with the un-encoded cases.

A direct application of our technique is ultra-fast quantum  communication. Moreover, for muti-party quantum communication, e.g. a quantum web conference, it is clear that the physical qubits overlaps are more frequent. Our technique can also be used to support more users compared with the conventional QDT scheme.

In summary, our scheme not only enlarges the applications of quantum teleportation or entanglement swapping as well as the QECC techniques, but also brings new thoughts to correct some nonlocal errors and improve other related technologies. Besides its applications in quantum communication field, some other potential ones may also deserve some attention.  One is  the temporal high-resolution for quantum signals from particular physical systems.  Some systems may allow us to directly or indirectly encode the signals to some auxiliary systems to form a QECC.  Then the related temporal resolution for the signals will be raised by our framework.  Another possible application may be used in quantum storage technology. If we store the quantum data on some physical systems with some interactions in between,  a similar scheme (store the information after encoding) can be proposed to protect the quantum information against errors more effectively. This may lead to a longer storage time, and it is no longer restricted to the isolated systems.

\begin{acknowledgments}
This work is supported by the NNSF of China (Grant No. 11405120) and the Fundamental Research Funds for the Central Universities.
\end{acknowledgments}

\section*{Appendix A. QEMS for some representative systems}

We consider some simplified cases with each of their systems containing several subsystems as shown in FIG. (\ref{fig1}) (a)-(d). From these representative examples, we can  see  easily that how QEMS works on these systems.
\begin{figure}
\begin{center}
 \includegraphics[scale=0.6]{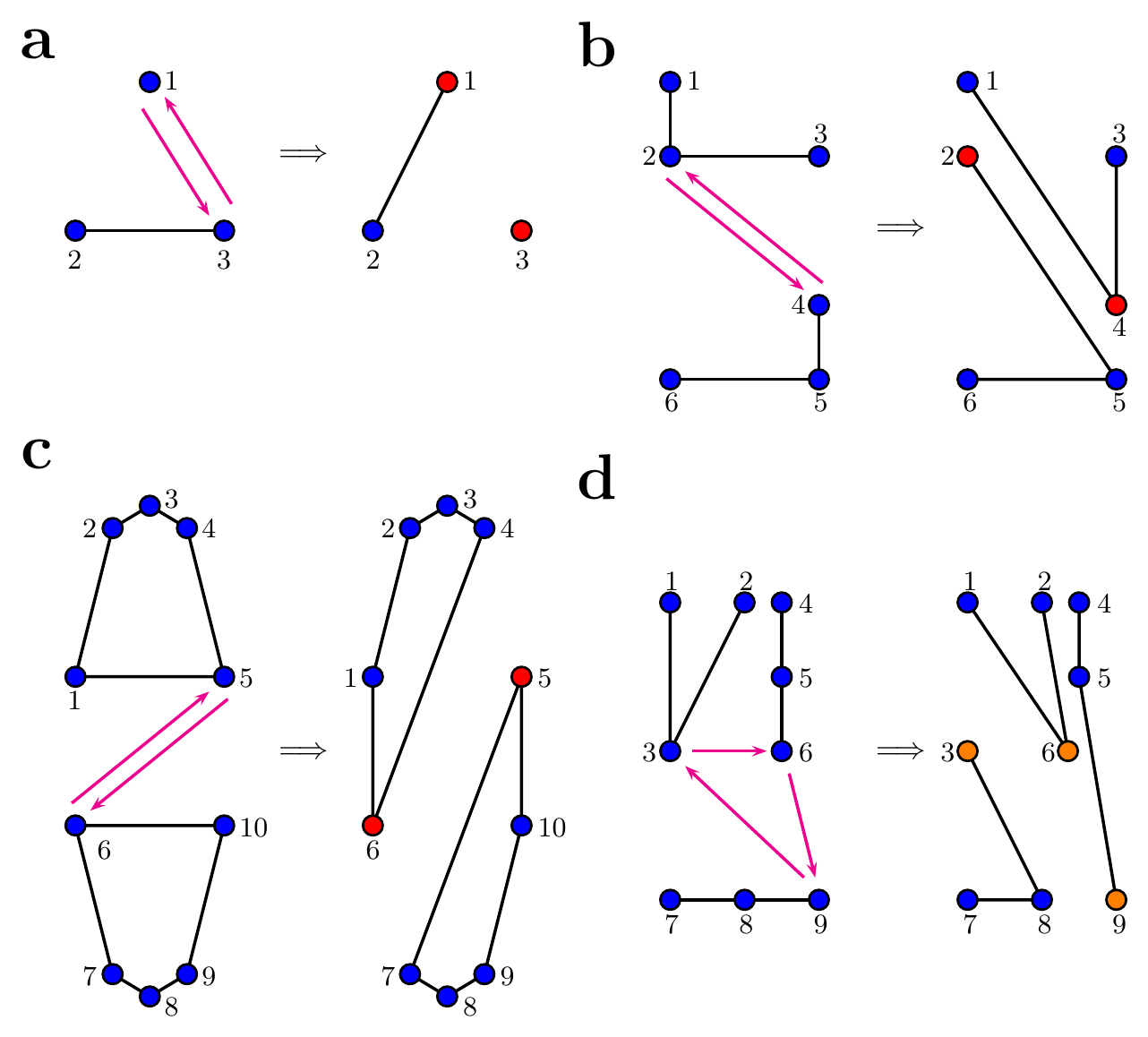}
\end{center}
\caption{\label{fig4}
{ QEMS from Graph states.}
({a}) A single qubit labeled by $1$ and a pair of Bell state labeled by $2$ and $3$. QEMS for qubits $1$ and $3$ can be revealed by a Bell measurement for qubits $1$ and $2$. This case is similar to the quantum teleportation scheme of an unknown qubit with the aid of a Bell state except for a extra unitary operation on qubit $3$ by the latter.
 ({b}) QEMS for qubits $2$ and $4$ from two GHZ state, which is essentially analogous to an entanglement swapping scheme for two GHZ state.
 ({c}) QEMS for qubits $2$ and $4$ from two five-qubit graph states.
 ({d}) Qubits permutation measurement statistics for three qubits from different GHZ states.}
\end{figure}

(a)Let $|\varphi\rangle$ be an qubit state and $|\Phi\rangle$ be one of the four common two-qubit Bell state, e.g. $\frac{1}{\sqrt{2}}(|00\rangle+|11\rangle)$. Clearly $|\varphi\rangle_1|\Phi\rangle_{23}=V_{13}|\Phi\rangle_{12}|\varphi\rangle_3=
\frac{1}{2}[|\Phi\rangle_{12}|\varphi\rangle_3+X_1|\Phi\rangle_{12}X_3|\varphi\rangle_3+
Y_1|\Phi\rangle_{12}Y_3|\varphi\rangle_3+Z_1|\Phi\rangle_{12}Z_3|\varphi\rangle_3]$.
Then a Bell measurement will be performed on the qubits $1$ and $2$,  and an equivalent error generated on the qubit $3$ can be inferred from QEMS. According to the outcomes from the Bell measurement, we can apply a unitary operation which is the same as the error occurring on the qubit $3$ to recover the imformation if we like.
It gives a more essential depiction for the expansion of a state from two different bases in the original quantum teleportation scheme from the framework of QEMS.

(b) Analogues to (a), we denote by $|G\rangle_{123}$ and $|G^{\prime}\rangle_{456}$ two  GHZ type states generated by qubits $1,2,3$ and qubits $4,5,6$, respectively. Then the  QEMS can be directly revealed from the expansion $|G\rangle_{123}|G^{\prime}\rangle_{456}=\frac{1}{2}[|G\rangle_{134}|G^{\prime}\rangle_{256}+
X_4|G\rangle_{134}X_2|G^{\prime}\rangle_{256}+Y_4|G\rangle_{134}Y_2|G^{\prime}\rangle_{256}
+Z_4|G\rangle_{134}Z_2|G^{\prime}\rangle_{256}]$.  Namely, if we measure the subsystem $1,3,4$ and $2,5,6$ by means of checking the parities of the stabilizers of $|G\rangle_{134}$ and $|G^{\prime}\rangle_{256}$, we will see a pair of ``twined errors" happens.

(c) Following the analysis of (b), likewise, QEMS for qubits $5$ and $6$ can be obtained by substituting $|G\rangle_{123}$ and $|G^{\prime}\rangle_{456}$ in (b) with two five-qubit graph states $|G\rangle_{1,2,3,4,5}$ and $|G^{\prime}\rangle_{6,7,8,9,10}$.

(d) This case  gives us a simple illustration of  $3$-qubit permutation measurement statistics which can be decomposed into two QEMS. Here on the qubits $3,6$ and $9$, apart from an possible global phase, the error here can only be one of the $16$ patterns: $I_iI_jI_k$ $(1)$, $X_iX_jI_k$ $(3)$, $Y_iY_jI_k$ $(3)$, $Z_iZ_jI_k$ $(3)$, and $X_iY_jZ_k$ $(6)$ after the measurements, where $(i,j,k)$ can be any permutation of $(3,6,9)$. Generally, for an $m$ qubits permutation case, the measurement statistics will give rise to any legal error type as a form of $\prod_{s=1}^m(\sigma_{\tau(s)})_s$ satisfying $\prod_{s=1}^m(\sigma_{\tau(s)})_k\propto I$ for any $k$, where $s$ denote the $s$-th qubit involved in the permutation and $\tau(s)\in\mZ_4$.

\section*{Appendix B. Some calculations for EDM and LDM}
To distinguish an event from which distribution optimally is usually related to a measure called the trace distance\cite{Nielsen}. By notice that,
we can roughly use the average area $S_c$  of  the common zone of two sequential distributions, e.g. the black zone in FIG.\ref{fig3}-(a) and (b), to link the probability $p$ for a QDE happening on each path, since a proportional relationship $p=\lambda S_c ~(\lambda>0)$ exists between them, and sometimes we may roughly write $p=\frac{1}{2}S_c$ due to the definition of trace distance.

As a physical qubit from signal A may wrongly appear in the time interval belonging to signal B (see FIG. \ref{fig1}-(c), the time interval between two dashed red lines), and the probability to characterize that is related to the integration of the tail part for the probability density function of the former signal A from $t_B$ to $t_B+l$ $(l=|t_A-t_B|)$. As a good approximation, if the convergence of the probability density function is rapid enough, we can integrate the function from $t_B$ to infinity. Even in the overlapped zone, there is still an opportunity to identify the two qubits correctly.  Therefore, the probability for a QDE happens between them  can only be regarded as a quantity proportional to $S_c$. Herein, we prefer to use $p=\frac{1}{2} S_c$ as it can provide us some numerical evaluation for the raising of the qubit rate limit.

(i)For the EDM in FIG.\ref{fig3}-(a), each probability density function can be described as
$f_i(t)=ae^{-a(t-t_{i})}, i=A,B, t\geq t_i$.  the probability for a QDE occurring, can be given by
\begin{align*}
    p=\frac{1}{2}\int_{l}^\infty dt ae^{-at}=\frac{1}{2}e^{-al}.
\end{align*}
Let $\frac{1}{2} e^{-aL}=q$, namely, $a=-\frac{\ln (2q)}{L}$, then we have $p=\frac{1}{2}(2q)^{\frac{l}{L}}$, which has already been used in the Results.

(ii)For the LDM in FIG.\ref{fig3}-(b), the relevant probability density function can be written as $f_i(t)=\sqrt{\frac{c}{2\pi}}\frac{e^{-\frac{c}{2(t-t_i)}}}{(t-t_i)^{\frac{3}{2}}}$. As the  integration of the probability density function mainly depends on its tail behavior, and we can simplify it to $f_i(t)=\sqrt{\frac{c}{2\pi}}\frac{1}{t^{\frac{3}{2}}}$ when $t$ is sufficiently large. Likewise,
\begin{align*}
    p=\frac{1}{2}\int_{l}^\infty dt \sqrt{\frac{c}{2\pi}}\frac{1}{t^{\frac{3}{2}}}=\frac{1}{2}\sqrt{\frac{2c}{\pi}}l^{-\frac{1}{2}},
\end{align*}
and $\frac{1}{2}\sqrt{\frac{2c}{\pi}}L^{-\frac{1}{2}}=q$, then we can get $p=\sqrt{\frac{L}{l}}q$. For a QECC of $[[n,1,2r+1]]_2$ type, we should consider the contributions for QDEs up to the order $p^{r+1}$. Analogous to the EDM's case in the Results,
we have
\begin{align*}
    (\frac{3}{4})^{r+1}C_n^{r+1}p^{r+1}\leq\frac{3}{4}q,
\end{align*}
then
\begin{align*}
    l\geq(\frac{3}{4})^2(\frac{4}{3})^{\frac{2}{r+1}}(C_n^{r+1})^{\frac{2}{r+1}}q^{-{\frac{2}{r+1}}}q^2L.
\end{align*}

For $r=1$, $l\geq\frac{3}{4}qC_n^2L$. If the pre-given $q=0.01$ and the code $[[5,1,3]]_2$ is chosen, we have $N_m^{\prime}=\frac{1}{l_{\min}}=\frac{4}{3}(C_5^2)^{-1}q^{-1}N_m\approx13.3N_m$.

But if $r$ is sufficiently large, the above inequality can be simplified to $l\geq(\frac{3}{4})^2q^2L$, namely, $N_m^{\prime}=\frac{1}{l_{\min}}\approx(\frac{4}{3})^2q^{-2}N_m$. E.g. for $q=0.01$, $N_m^{\prime}\approx17778N_m$,
which indicates that the qubit rate can be raised by 4 orders of magnitude!
However, this calculation is based on the approximate relationship $C_n^{r+1}p^{r+1}(1-p)^{n-r-1}\approx C_n^{r+1}p^{r+1}$ and in the zone where the probability density function can be approximately described by its tail behavior. So generally, if the code $[[n,1,2r+1]]_2$ we chosen satisfied $2r+1<20$, by a conservative estimate, raising the qubit rate by one or two orders of magnitude may be easily achieved. But if the whole device requires a much smaller $q$, e.g. $q\leq0.001$, the qubit rate limit will be raised by  more than two orders of magnitude.

More generally, if the asymptotic tail behavior of the probability density function goes like $\alpha x^{-s}$ with $\alpha>0$ and $s>1$. We can get $p=(\frac{L}{l})^{s-1}q$, and $N_m^{\prime}=(\frac{4}{3})^{\frac{1}{s-1}}q^{-\frac{1}{s-1}}\cdot(\frac{3}{4})^{\frac{1}{(r+1)(s-1)}}
q^{\frac{1}{(r+1)(s-1)}}(C_n^{r+1})^{-\frac{1}{(r+1)(s-1)}}\cdot N_m$ the QECC $[[n,1,2r+1]]_2$.
For the case $(r+1)(s-1)\gg1$, the above formula will be $N_m^{\prime}\simeq(\frac{4}{3})^{\frac{1}{s-1}}q^{-\frac{1}{s-1}}N_m$.
For example, if $s=2, r=1, q=0.01, n=5$, then $N_m^{\prime}\approx3.65N_m$. When a large $r$ is chosen, $N_m^{\prime}\sim133N_m$.

In brief, the improvement for the qubit rate mainly depends on the asymptotic tail behavior of the relevant probability density function. For some common distributions, by choosing a proper QECC, we can raise the qubit rate by several times to hundreds of times, which seems a significant improvement.

\section*{Appendix C. Extra remarks on the qubit rate upper bound for EDM}
For a general $r$ case, we can not ensure that $(2q)^{\frac{1}{r+1}}$ is small enough. Suppose that $C_n^{r+2}p^{r+2}(1-p)^{n-r-2}\ll C_n^{r+1}p^{r+1}(1-p)^{n-r-1}$ still holds, i.e., $p\ll\frac{r+2}{n+1}$ or the rate $N_m^{\prime}\ll\frac{N_m}{\log_{2q}\frac{2(r+2)}{n+1}}$, then we consider the QDEs up to order $p^{r+1}$(the number of errors $\leq r$ can be corrected) and we should ensure
\begin{align}\label{exact}
 (\frac{3}{4})^{r+1}C_n^{r+1}p^{r+1}(1-p)^{n-(r+1)}\leq\frac{3}{4}q,
\end{align}
where $p=\frac{1}{2}(2q)^{\frac{N_m}{N_m^{\prime}}}$. As $(1-p)^{n-(r+1)}<1$, thus the rate $N_m^{\prime}$ is smaller than $(r+1)N_m$. For any given $q$, $n$ and $r$ we can solve Eq. (\ref{exact}) numerically and get the corresponding $N_m^{\prime}$. In all, for a general integer $r$, $N_m^{\prime}$ no longer increases linearly with it.


\begin{thebibliography}{99}
\bibitem{highorder} Actually, three or more pairs of QDEs should also be considerred. However, we can consider the case when $p$ is not sufficiently small but still not too large, if $C_n^2p^2(1-p)^{n-2}\gg C_n^3p^3(1-p)^{n-3}$, i.e., $p\ll\frac{3}{n+1}$, then we can ignore the contribution from the three QDEs. Besides, the contribution for the cases of one physical qubit involved in two or more QDEs can also be neglected, as these cases can also produce single error on this qubit accroding to the form of qubits permutation operators.
\end{thebibliography}
\end{document}